\documentclass[10pt, conference]{IEEEtran}
\IEEEoverridecommandlockouts

\usepackage{cite}
\usepackage{amsmath,amssymb,amsfonts}
\usepackage{algorithmic}
\usepackage{graphicx}
\usepackage{textcomp}
\usepackage{xcolor}
\usepackage{multirow}
\usepackage{url}
\usepackage{balance}
\usepackage{tcolorbox}
\def\BibTeX{{\rm B\kern-.05em{\sc i\kern-.025em b}\kern-.08em
    T\kern-.1667em\lower.7ex\hbox{E}\kern-.125emX}}
\begin{document}

\title{Apples, Oranges \& Fruits - Understanding Similarity of Software Projects Through The Lens of Dissimilar Artifacts}

\author{\IEEEauthorblockN{A Eashaan Rao and Sridhar Chimalakonda}
\IEEEauthorblockA{\textit{Research in Intelligent Software \& Human Analytics (RISHA) Lab}\\
Department of Computer Science \& Engineering \\
Indian Institute of Technology Tirupati India\\
\{cs19s501,ch\}@iittp.ac.in}
}

\maketitle

\begin{abstract}

The growing availability of open source projects has facilitated developers to reuse existing software artifacts and leverage them to develop new software. However, it is hard to understand the notion of similarity as it varies from developer to developer. Some developers might search for repositories with similar source code, while some might be in search of repositories with similar requirements or issues. Existing approaches tend to find similar projects by \textit{comparing similar artifacts} such as source-code to source-code, API usage to API usage, documentation to documentation, and so on. Even though there is a dissimilarity between two similar artifacts, there could be a similarity between two dissimilar artifacts. Hence, in this paper, we aim to answer the question - \textit{Can we find similarity of software repositories through dissimilar artifacts?}. To this end, we conduct an experiment to find similarities between three repositories, two similar and one different project comparing similar and dissimilar artifacts (documentation, commits, and source-code). We observed similarities between dissimilar artifacts such as \textit{Commits}, \textit{Source Code}, and \textit{Readme Files} in the context of both similar and different repositories.

\end{abstract}

\begin{IEEEkeywords}
Software Similarity, Software Artifacts, Documentation, Commits, Source-Code
\end{IEEEkeywords}

\section{Introduction}
\label{sec:Intro}
Open-Source Development (OSD) and code hosting platforms such as GitHub facilitate developers to study and understand techniques used in different repositories\footnote{Projects and Repositories are interchangeably used for the term Software Repositories} \cite{nafi2018crolsim}. High-quality open-source repositories provide a wealth of information in terms of software implementation and its maintenance \cite{nguyen2020automated}. Therefore, if a developer wants to build a software system, they may refer to similar and ``well-defined, mature projects'' \cite{nguyen2020automated} to understand the development activities, and to improve the software quality \cite{capiluppi2020detecting}. Hence, finding similar repositories from the myriad of repositories is essential for developers to reduce their time-effort in developing new software \cite{zhang2017detecting}. 

Zhang et al. \cite{zhang2017detecting} defines that two repositories are similar if they perform similar functionalities or have similar source code files. Mcmillan et al. \cite{mcmillan2012detecting} state that two software projects are similar if the same requirements are described by similar functional abstractions.  Kawaguchi et al. \cite{kawaguchi2006mudablue} consider similar software based on the identical identifiers found in the software systems. Evidently, there is no precise definition for similar repositories as it varies from the way they are extracted  \cite{nguyen2020automated,capiluppi2020detecting}, as well as the purpose of finding similar repositories, may vary for developers \cite{zhang2017detecting,nguyen2019focus,xia2017developers}. 

There are many tools and techniques proposed to find similar software projects, such as MUDABlue \cite{kawaguchi2006mudablue} creates a categorization of similar software systems using LSA, CLAN \cite{mcmillan2012detecting} utilizes API calls to find similar functional abstraction which implements similar requirements for different repositories. RepoPal \cite{zhang2017detecting} used high-level information, such as, readme files and stargazers details
to calculate relevance score between two repositories to find similarity. Capiluppi et al. \cite{capiluppi2020detecting} used CrossSim \cite{nguyen2020automated, nguyen2018crosssim} to analyze the software systems by forming clusters of similar software and applied object-oriented metrics on it to differentiate the clusters. 

The techniques presented in the literature majorly focuses on two types of similarity computation, i.e., low-level similarity (comparison among source-code, functions, API calls) and high-level similarity (comparison among project's metadata such as documentation and stargazers) \cite{zhang2017detecting,mcmillan2012detecting,capiluppi2020detecting, chen2015simapp}. These techniques are used in clustering to categorize software into groups \cite{mcmillan2012detecting,capiluppi2020detecting}. However,\textbf{\textit{ due to the presence of interconnectedness among the artifacts, a cross-relationship exists among them, which has not been utilized to find similar repositories}} \cite{nguyen2020automated}. For example, documentation files are related to the source code, which itself is related to test cases, issues, architecture, and APIs. Despite the affinity present among different software artifacts, they are not used as a comparison technique in the context of finding similar software repositories.

In this paper, we propose that the similarity between two repositories can also be found by comparing two dissimilar artifacts. For instance, Artifact X from one repository can be taken as input and can be compared against Artifact Y from other repositories. For instance, documentation present in repository A can be used as input to find relevant source-code from repository B. This concept is utilized in research areas of code-search \cite{poshyvanyk2007combining,poshyvanyk2013concept,gu2018deep, sivaraman2019active}, bug-localization \cite{zhou2012should,lam2017bug,khatiwada2020combining}, and so on. However, this aspect has been overlooked in finding similar software repositories.\textit{ The core argument of this paper is to highlight the need to look at dissimilar artifacts as comparison criteria to discover similar software repositories.} The necessity to broaden the prospects for finding similar software repositories by utilizing dissimilar software artifacts might provide flexibility in searching and suggesting similar repositories \cite{nguyen2020automated}.

\section{Motivation Scenario and Hypothesis}
According to the literature, detecting similar repositories helps the developers to look at alternative implementations and learn about the ``best practices and programming idioms,'' \cite{kawaguchi2006mudablue} which improves the understanding of source code, rapid prototyping, promote code reuse, and helps to discover plagiarism cases \cite{zhang2017detecting, mcmillan2012detecting, capiluppi2020detecting}. However, the fundamental challenge in detecting similar repositories is the mismatch of high-level description and lower-level implementation \cite{mcmillan2012detecting}.

Even though significant research is done to find similar repositories, there are still noteworthy limitations present in the proposed tools/techniques. MUDABlue \cite{kawaguchi2006mudablue} automatically categorizes software by extracting identifiers and using LSA to retrieve latent relation between them; hence these identifiers can act as a tag to classify software. However, some of the extracted identifiers are not easy to understand \cite{kawaguchi2006mudablue}. Software tags commonly highlight the domain and are not descriptive enough, also they are not found in many GitHub repositories \cite{zhang2017detecting}. CLAN \cite{mcmillan2012detecting} uses API calls to extract the semantic relationship between repositories; however, its core idea of finding similar abstractions between repositories in a dynamic environment (frequent updates) such as in GitHub is a cumbersome task \cite{zhang2017detecting}. RepoPal weighs three heuristics to detect similar repositories, i.e., similar users, readme files, and the short duration, in which the same user stars multiple repositories \cite{zhang2017detecting}. Still, one can argue in this approach that developers who are looking at specific repositories might get different results due to the stargazer relevance and time-relevance score, as selecting repositories based on highest stars may have resulted due to the possibly active social media promotion \cite{borges2018s}, rather than the similar interests. 

The common limitation in the approaches mentioned above is that they work only in a limited scope and do not consider additional metadata to compute similarity \cite{nguyen2020automated}. To tackle this, CrossSim \cite{nguyen2018crosssim,nguyen2020automated} creates RDF graphs for software to make use of different relationships between various ``actors,'' such as developers, source code, libraries, commits, and repository stars. The comparison between graphs is made to compute similarity among various nodes using several graph algorithms. Regardless of combining multiple artifacts under one roof in the form of the graphs to facilitate similarity computation, the formation of a graph and the graph structure (how multiple artifacts are placed together, are they making full use of the data present in the artifacts, and so on) itself influences the CrossSim performance \cite{nguyen2020automated}. 

The common thread between the techniques mentioned above is that they compare the same types of artifacts. MUDABlue compares source code, CLAN compares API call usages, RepoPal uses readme files, and CrossSim compares graphs. As discussed in Section \ref{sec:Intro}, the relationship between dissimilar artifacts is not explored in this context.

\subsection{Hypothesis}

This paper argues that there are multiple correlations that exist between artifacts such as documentation, requirements, source code, design diagrams, architecture, test cases, and reported issues that are yet to be explored in the context of finding similar repositories. There is a need to broaden the scope in defining similar repositories and the methodology for searching them where a correlation can be establish between dissimilar artifacts using software traceability techniques \cite{sulun2021rstrace+}.

There are numerous scenarios where dissimilar artifacts are linked together to solve multiple software engineering problems. In code search, a natural text is taken as input to produce related source-code snippets as output \cite{poshyvanyk2007combining}. Cross-project defect prediction uses the defect history of one software project to predict probable defective areas of another software project \cite{herbold2017global}.  In bug localization,  bug reports and fixed source code snippets are taken into consideration to predict the origin of the bug mentioned in a newly reported bug report \cite{lam2017bug}. Requirement traceability links the requirements with source code as well as test cases \cite{ali2019exploiting}. Mutation testing uses the idea of introducing artificial test defects in source code to support testing activities \cite{grano2019lightweight}.

Furthermore, we also have to contemplate on how developers use search engines to find similar software. In a few instances, they might not be sure about what exactly they are searching for in similar repositories, or maybe they are searching for some concepts/techniques used in different applications/domains \cite{stolee2014solving, martie2017understanding}. Once they get the search results, they might not be sure whether the artifacts they found in a repository are the ones they need. Hence it will become an iterative process of searching \cite{martie2017understanding}. The more time they spend on querying and looking at the results, the more clearer their query will become \cite{martie2017understanding}. Developers may want to find some specific implementation for a particular requirement, not for the whole system \cite{xia2017developers}; for them, the definition of a similar repository may be restricted to correlated requirements and source code.  Besides, the current techniques/approaches may not able to recommend such results which may requires inclusion of dissimilar artifacts, since it compares only similar artifacts. Our next step in this research area should move towards uncovering the relations among various artifacts to compute software similarity from various points of view. In addition to it, applying the tools and techniques used in other research areas such as defect prediction, code search, bug localization, and so on, and tailoring them in the context of finding similar software repositories might help us to utilize multiple artifacts.

\begin{table*}[]
\centering
\caption{Similarity among Similar and Dissimilar Artifacts for Related and Unrelated Repositories}
\label{tab:results}
\begin{tabular}{c|l|l|l|c|}
\cline{2-5}
\multicolumn{1}{l|}{}                                                                                                       & \textbf{Vectorizing Method}        & \textbf{Repository\_Name}                                                                                & \textbf{Arifacts}                       & \multicolumn{1}{l|}{\textbf{Highest Cosine Similarity}} \\ \hline
\multicolumn{1}{|c|}{\multirow{8}{*}{\textbf{\begin{tabular}[c]{@{}c@{}}Comparing Similar \\ Repositories\end{tabular}}}}   & \multirow{4}{*}{Tf-idf Vectorizer} & \multirow{4}{*}{\begin{tabular}[c]{@{}l@{}}JamsMusicPlayer (a)\\ \\ Vk\_music\_android (b)\end{tabular}} & Source Code (a) Vs Source Code (b)      & 0.945                                                   \\ \cline{4-5} 
\multicolumn{1}{|c|}{}                                                                                                      &                                    &                                                                                                          & Commits (a) Vs Commits (b)              & 0.726                                                   \\ \cline{4-5} 
\multicolumn{1}{|c|}{}                                                                                                      &                                    &                                                                                                          & \textbf{Commits (a) Vs Source Code (b)} & \textbf{0.415}                                          \\ \cline{4-5} 
\multicolumn{1}{|c|}{}                                                                                                      &                                    &                                                                                                          & \textbf{Readme (a) Vs Source Code (b)}  & \textbf{0.535}                                          \\ \cline{2-5} 
\multicolumn{1}{|c|}{}                                                                                                      & \multirow{4}{*}{CountVectorizer}   & \multirow{4}{*}{\begin{tabular}[c]{@{}l@{}}JamsMusicPlayer (a)\\ \\ Vk\_music\_android (b)\end{tabular}} & Source Code (a) Vs Source Code (b)      & 0.97                                                    \\ \cline{4-5} 
\multicolumn{1}{|c|}{}                                                                                                      &                                    &                                                                                                          & Commits (a) Vs Commits (b)              & 0.839                                                   \\ \cline{4-5} 
\multicolumn{1}{|c|}{}                                                                                                      &                                    &                                                                                                          & \textbf{Commits (a) Vs Source Code (b)} & \textbf{0.569}                                          \\ \cline{4-5} 
\multicolumn{1}{|c|}{}                                                                                                      &                                    &                                                                                                          & \textbf{Readme (a) Vs Source Code (b)}  & \textbf{0.673}                                          \\ \hline
\multicolumn{1}{|c|}{\multirow{8}{*}{\textbf{\begin{tabular}[c]{@{}c@{}}Comparing Different \\ Repositories\end{tabular}}}} & \multirow{4}{*}{Tf-idf Vectorizer} & \multirow{4}{*}{\begin{tabular}[c]{@{}l@{}}JamsMusicPlayer (a)\\ \\ Lighting-Browser (b)\end{tabular}}   & Source Code (a) Vs Source Code (b)      & 0.963                                                   \\ \cline{4-5} 
\multicolumn{1}{|c|}{}                                                                                                      &                                    &                                                                                                          & Commits (a) Vs Commits (b)              & 0.692                                                   \\ \cline{4-5} 
\multicolumn{1}{|c|}{}                                                                                                      &                                    &                                                                                                          & \textbf{Commits (a) Vs Source Code (b)} & \textbf{0.166}                                          \\ \cline{4-5} 
\multicolumn{1}{|c|}{}                                                                                                      &                                    &                                                                                                          & \textbf{Readme (a) Vs Source Code (b)}  & \textbf{0.268}                                          \\ \cline{2-5} 
\multicolumn{1}{|c|}{}                                                                                                      & \multirow{4}{*}{CountVectorizer}   & \multirow{4}{*}{\begin{tabular}[c]{@{}l@{}}JamsMusicPlayer (a)\\ \\ Lighting-Browser (b)\end{tabular}}   & Source Code (a) Vs Source Code (b)      & 0.977                                                   \\ \cline{4-5} 
\multicolumn{1}{|c|}{}                                                                                                      &                                    &                                                                                                          & Commits (a) Vs Commits (b)              & 0.814                                                   \\ \cline{4-5} 
\multicolumn{1}{|c|}{}                                                                                                      &                                    &                                                                                                          & \textbf{Commits (a) Vs Source Code (b)} & \textbf{0.273}                                          \\ \cline{4-5} 
\multicolumn{1}{|c|}{}                                                                                                      &                                    &                                                                                                          & \textbf{Readme (a) Vs Source Code (b)}  & \textbf{0.382}                                          \\ \hline
\end{tabular}
\end{table*}

\section{Proposed Approach}
\label{sec:approach}
\begin{figure}
    \centering
    \includegraphics[scale=0.44]{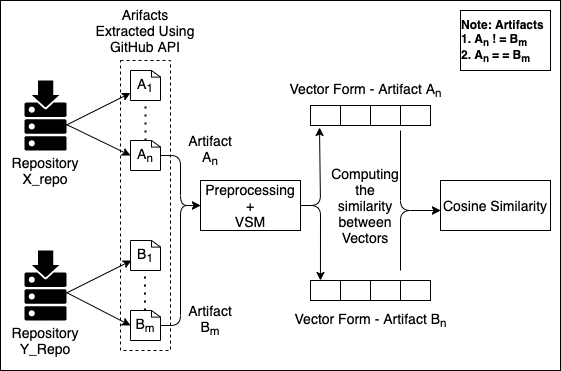}
    \caption{Proposed Approach for finding Textual Similarity in Similar \& Dissimilar Artifacts}
    \label{fig:Dis_Artifacts_Approach}
\end{figure}

In this section, we are going to discuss one of the approaches that can be used to compare two dissimilar artifacts from two different repositories. The purpose of these experiments is to try to establish that given two similar/different repositories, there is an association that exists between dissimilar artifacts, and we can leverage this association to find similar repositories. Fig \ref{fig:Dis_Artifacts_Approach} describes the proposed approach and to find the links among two dissimilar artifacts, say A and B, we carry out the following strategy:
\begin{itemize}
    \item Artifacts considered for the experiments: Source Code, Commits Data, and Readme files, which are extracted using GitHub API (see Fig \ref{fig:Dis_Artifacts_Approach}).
    \item We considered a \textit{Base\_Repository} say \textit{X\_repo} and based on its domain, we found a similar repository. For example, if \textit{X\_repo} is a photo application, we will consider a similar photo application built on the same framework, say \textit{Y\_repo}.
    \item We consider another repository \textit{Z\_repo} whose functionality is different from \textit{X\_repo}, for example, health tracking application. For this study, we made an assumption that the developer might be looking at the repositories which are built on a similar framework.
    \item From \textit{X\_repo}, we choose artifact A, and from \textit{Y\_repo} or \textit{Z\_repo}, we choose artifact B (see Fig \ref{fig:Dis_Artifacts_Approach}). To discover the connection among artifacts A and B, we calculate the textual similarity between them using a \textit{vector space model (VSM)}. VSM converts the artifacts into vectors of identifiers. Cosine similarity is then employed to determine the similarity between these vectors. This technique is used in many research areas such as bug localization, to find relevant source files for a given bug report using textual similarity\cite{zhou2012should}.
    \item Artifacts are preprocessed before converting into vectors. Preprocessing steps include tokenization, removing stop words, and stemming. Conversion into a vector is accomplished using two functions Tfidfvectorizer and CountVectorizer from the sklearn library\footnote{\url{https://scikit-learn.org/stable/modules/classes}}. Tf-idf function weighs an identifier presence in a corpus by calculating identifier's frequency in a document and in the corpus. CountVectorizer counts the word frequencies in the document. We selected both techniques because there might be some cases where certain identifiers occurs more frequently; however, due to the tf-idf, its value lowered, whereas CountVectorizer might give appropriate weight to that identifier. We calculate the textual similarity score between artifacts by considering both vectorizing approaches.
\end{itemize}

\section{Experiments}
\label{sec:expt}
We selected three repositories, namely, \textit{JamsMusicPlayer}\footnote{\url{https://github.com/psaravan/JamsMusicPlayer}} and  \textit{Vk\_music\_android} \footnote{\url{https://github.com/Mavamaarten/vk_music_android}}, which are android-based music applications taken from the \textit{CrossSim}\footnote{\url{https://github.com/crossminer/CrossSim}} dataset \cite{nguyen2018crosssim}, and \textit{Lighting-Browser}\footnote{\url{https://github.com/anthonycr/Lightning-Browser}}, a lightweight android browser application. Based on the assumption, as described in Section \ref{sec:approach}, the repositories considered for this study are built on Android framework.  We consider \textit{JamsMusicPlayer} as a base repository. As mentioned in Section \ref{sec:approach}, we selected three artifacts for comparison, i.e., source code, commits, and readme files. Experiments consist of two parts: comparing similar repositories and comparing different repositories based on similar and dissimilar artifacts. We are finding the textual similarity based on different combinations of artifacts, i.e., \textit{Source\_Code Vs. Commits\_Data}, \textit{Readme\_file Vs. Source\_Code} (dissimilar artifacts), and \textit{Source\_Code Vs. Source\_Code}, \textit{Commits\_Data Vs. Commit\_Data} (similar artifacts). Comparing textual similarity in similar and dissimilar artifacts will give us an idea about the nature of results, which will help us understand the challenges when comparing dissimilar artifacts.

For example, \textit{Commits\_Data}, is taken from \textit{JamsMusicPlayer} and \textit{Source\_Code} from \textit{Vk\_music\_android}. The \textit{Commits\_Data} contains all the commits in a text file, and all the \textit{Source\_Code} files are stored in a single folder. Using the pipeline, as shown in Fig \ref{fig:Dis_Artifacts_Approach}, \textit{Commits\_Data} will be compared with each \textit{Source\_Code} file and calculates the cosine similarity between them, similar process is applied between \textit{ReadMe\_file} Vs. \textit{Source\_Code}. When comparing similar artifacts, such as \textit{Source\_Code}, each file from \textit{X\_repo} is compared with each file from \textit{Y\_repo} or \textit{Z\_repo}. 

\section{Preliminary Analysis}
Table \ref{tab:results} presents the results of the experiments discussed in Section \ref{sec:expt}. We show that the highest cosine similarity is achieved while comparing multiple dissimilar artifacts. We have used two types of vectorizers, as described in Section \ref{sec:approach}. We observed from the results that when similar artifacts are compared irrespective of whether the repositories are similar or different, the cosine similarity between them is at least 70\%. The rationale behind this observation is that all the applications considered for the experiments are developed in a similar framework, i.e., Android. Hence, the majority of the files may have the same kind of structure and are relatable to a certain extent.

For dissimilar artifacts, the cosine similarity value is less than 50\% in most cases. Still, the numbers are promising, as these similarity values are over 50\% in some cases for similar repositories. Even though cosine values are low in the context of different repositories, i.e., approx. 20-30\%, yet this confirms our hypothesis that similarity does exist between dissimilar artifacts to a certain extent, in both the similar and different repositories context, which could be an emerging result. In a few cases, the value of cosine similarity is found to be zero.

In the context of the vectorizing methods, \textit{CountVectorizer} has shown high similarity values than \textit{Tf-idf Vectorizer}. However, we also have to consider the nature of the vectorizing techniques, as described in Section \ref{sec:approach}. \textit{Tf-idf} is more restrictive towards frequent identifiers, which we assume that in the context of source code may not be applicable. Therefore, we compare \textit{Tf-idf} with \textit{CountVectorizer} to know the difference between the output results and found that the difference is not statistically significant.

\begin{tcolorbox}
Textual similarity exists between dissimilar artifacts such as Commits, Source\_Code, and Readme\_Files, in the context of similar and different repositories.
\end{tcolorbox}

\section{Discussion and Challenges}

We used the title of the paper ``Apples, Oranges \& Fruits,'' to imply that even though Apples and Oranges are different, they still belong to the category of fruits. Therefore, they are similar and related to each other to a certain extent from a different perspective. Equivalently, we also try to show that dissimilar artifacts such as Source\_Code and Commit\_Data are also related to each other by finding the textual similarity between them. We employed a \textit{vector space model (VSM)}, which is a basic technique to compute the textual similarity between the artifacts. However, using the techniques from Natural Language Processing (NLP), Machine-Learning (ML), or Deep-Learning (DL) Models in the future, we might establish more concrete links between these dissimilar artifacts by leveraging their lexical and semantic relationship. The scope of this study is limited to define a relationship between dissimilar artifacts. However, producing a ranked list of similar repositories based on a query is our primary goal, which will be undertaken in future studies. Comparing dissimilar artifacts from two perspectives, as mentioned in Section \ref{sec:expt} allows us to understand that a repository can be related to another repository to a certain extent. 

One can argue that matching similar words from the artifacts may lead to a lexical mismatch and may not guarantee that two applications are similar \cite{mcmillan2012detecting}. However, due to the recent advancements in \textit{NLP, ML, and DL}, underlying relationships among artifacts can be found by considering their structural and semantic information \cite{huo2019deep,wang2018deep,allamanis2018survey}, which gives us the confidence that a concrete association can also be found between dissimilar artifacts. One of the challenges that we might have to consider is the quality of the artifacts. High-quality artifacts can be found in well-matured repositories \cite{nguyen2020automated},  yet many repositories are not well-maintained \cite{coelho2020github}.  Also, a new developer may not differentiate between good/bad quality artifacts. Therefore, finding high-quality similar repositories is one of the challenges that we need to take into account. We also need to look at how developers search for similar repositories. For example, it might be the natural text of software tags or the framework used \cite{xia2017developers}. Additionally, we have to examine the specific artifacts that a developer might be looking for in similar repositories, such as code snippets, architecture design, test cases, issues, or project structure. Therefore, we believe that comparing dissimilar artifacts is one of the approaches that could satisfy such use cases.

\section{Conclusion and Future Work}
The paper argues that finding similarity between software repositories is a complex task, and the proposed techniques in the literature do not focus on finding similar software repositories from multiple perspectives, and there is a need to shift our focus to dissimilar artifacts. We showed that associations does exist between dissimilar artifacts by computing the textual similarity between \textit{Commits\_data}, \textit{Source\_Code} and \textit{Readme\_files} for similar and different repositories. Dissimilar artifacts are found to be 40-50\% similar in the context of similar repositories and 20-30\% for dissimilar repositories. 

We see this paper as a promising direction for further research towards finding similarity among dissimilar artifacts. An important future work is to apply various NLP/ML/DL techniques to compute similarity among dissimilar artifacts. We are planning to present similar repositories results from multiple perspectives, i.e., with respect to different artifacts, domain, language, framework, and so on. We are also focusing on the query reformulation for finding similar repositories.

\balance
\bibliographystyle{IEEEtran}
\bibliography{dissimilar_artifacts}

\end{document}